\documentclass[pra,aps]{revtex4}
\usepackage{graphicx}
\usepackage{epsfig}
\newtheorem{theorem}{Theorem}

\def\numbit{n}
\def\numdim{N}
\def\uunpert{U}
\def\upert{U'}
\def\hs{{\cal H}}
\def\coin{C}
\def\grover{G}
\def\tensorm{\otimes}
\def\shift{S}
\def\dotm{\cdot}
\def\xhc{\vec{x}}
\def\khc{\vec{k}}
\def\kl{k}
\def\dhc{d}
\def\dir{e}
\def\modp{\oplus}
\def\ket#1{\left\vert #1 \right\rangle}
\def\bra#1{\left\langle #1 \right\vert}
\def\bkm#1#2{\left\langle #1 \vert #2 \right\rangle}
\def\id{{\cal I}}
\def\eqsupc{s^{\coin}}
\def\eqsups{s^{\shift}}
\def\eigunpert{\omega}
\def\eigunpertvec{v}
\def\eigpert{\eigunpert'}

\def\bigo{O}

\def\by{\times}
\def\e{e}
\def\coinpert{\coin'}
\def\inita{\psi_0}
\def\initm{\psi_-}
\def\initp{\psi_+}
\def\initb{\psi_1}

\def\norminitb{c}
\def\swapij{P_{ij}}
\def\time{t}

\def\dist#1{\left\vert #1 \right\vert}
\def\xl{x}
\def\nchoosek#1#2{{#1 \choose #2}}
\def\abs#1{\left\vert #1 \right\vert}

\def\eigtrial{\alpha}

\def\arcimp{{\cal A}}

\def\eigimp{\tilde\eigpert}
\def\commutator#1#2{\left[#1, #2\right]}

\def\prob{p}

\def\conj#1{#1^*}
\def\abphase{\eta}
\def\res{r}
\def\resb{r'}
\def\eigimpres{\beta}

\def\coeff{c}
\def\coeffb{c'}
\def\upcoeff{b}
\def\func{f}
\def\subspaceimp{\Omega}
\def\trace{\mathrm{Tr }}
\def\real{\mathrm{Re\:}}
\def\imag{\mathrm{Im\:}}
\def\xhctarget{\xhc_{target}}

\def\refsec#1{Sec.\ \ref{Section::#1}}

\def\reffig#1{Fig.\ \ref{Figure::#1}}

\def\refeqn#1{Eq.\ (\ref{Equation::#1})}
\def\refthm#1{Theorem \ (\ref{Theorem::#1})}

\def\refeqs#1#2{Eqs.\ (\ref{Equation::#1}) and (\ref{Equation::#2})}

\def\bbox{\vrule height7pt width4pt depth1pt}

\begin{document}


\title{A Quantum Random Walk Search Algorithm}

\author{Neil Shenvi$^1$}
\author{Julia Kempe$^{1,2,3}$}
\author{K. Birgitta Whaley$^1$}
\affiliation{Departments of Chemistry$^1$ and Computer
Science$^2$, University of California, Berkeley, CA 94720\\
$^3$CNRS-LRI, UMR 8623, Universit\'e de Paris-Sud, 91405 Orsay, France }

\date{\today}

\begin{abstract}
Quantum random walks on graphs have been shown to display many
interesting properties, including exponentially fast hitting times
when
compared with their classical counterparts.  However, it is still
unclear
how to use these novel properties to gain an algorithmic speed-up
over
classical algorithms.  In this paper, we present a quantum search
algorithm based on the quantum random walk architecture that
provides such
a speed-up.  It will be shown
that this algorithm performs an oracle search on a database of
$\numdim$
items with $\bigo\left(\sqrt{\numdim}\right)$ calls to the oracle,
yielding a
speed-up similar to other quantum search algorithms.  It appears
that the quantum random walk formulation has considerable
flexibility, presenting interesting opportunities for development
of
other, possibly novel quantum algorithms.
\end{abstract}
\pacs{}

\maketitle

\setcounter{figure}{0}

\section{Introduction} \label{Section::Introduction}
Recent studies of quantum random walks have suggested that they may
display different behavior than their classical counterparts
\cite{Aharonov:93a,Aharonov:01,Ambainis:01,Farhi:97,Childs:01}.  One
of the promising features of these quantum random walks is that they
provide an intuitive framework on which to build novel quantum
algorithms.  Since many classical algorithms can be formulated in
terms of random walks, it is hoped that some of these may be
translated into quantum algorithms which run faster than their
classical counterparts.  However, previous to a very recent paper by Childs et. al. 
\cite{Childs:02}, there had been no quantum algorithms based on the random walk model. 
In this paper we show that a quantum search algorithm can be derived from a certain
kind of quantum random walk.  Optimal quantum search algorithms are
already well known\cite{Grover:96,Grover:97,Bennett:97}. The search
algorithm from a quantum random walk we present here shows some
differences from the established search algorithms and may possess
useful properties with respect to robustness to noise and ease of
physical implementation.  It also provides a new direction for design
of quantum algorithms from random walks, which may eventually lead to
entirely new algorithms. 

Current research uses two distinct models for quantum random walks,
based on either discrete time steps or on continuous time evolution.
Discrete time quantum random walks were introduced as a possible new
tool for quantum algorithms generalizing discrete classical Markov
chains \cite{Aharonov:01}.  The discrete time walk can be thought of
as a succession of unitary operations, each of which has a non-zero
transition amplitude only between neighboring nodes of the graph.  The
relation of these to classical Markov chains provides considerable
motivation for exploration of discrete random walks.  Within the field
of classical algorithms, the application of classical Markov chains in
{\em classical} algorithms has been quite revolutionary, providing new
approximation and optimization algorithms. By analogy, it might
reasonably be hoped that similar algorithmic advances could be
obtained for quantum algorithms from development of the quantum random
walks.  The second quantum random walk model is the continuous-time
quantum random walk, introduced in \cite{Farhi:97,Childs:01,Childs:02}.  In the
continuous-time walk the adjacency matrix of the graph is used to
construct a Hamiltonian which gives rise to a continuous time
evolution. This model differs from the discrete time walk in that even
for small times there is an (exponentially small) probability of
transition to non-adjacent nodes. 
In this paper, we will consider the discrete-time
model only.

The paper is organized as follows.  \refsec{Background} provides a
brief introduction to discrete-time quantum random walks.
\refsec{RWSA} describes the random walk search algorithm and provides
a proof of its correctness.  \refsec{Grover} summarizes the
similarities and differences between the random walk search algorithm
and Grover's search algorithm.  Conclusions are presented in
\refsec{Conclusions}.

{\em Notation:} Following standard computer science notation we will use the following to characterize the growth of certain functions: We will say $f(n)=\bigo ( g(n))$ if there are positive  constants $c$ and $k$ such that $0 \leq f(n) \leq cg(n)$ for $n \geq k$. Similarly $f(n)=\Omega ( g(n))$ if $0 \leq c g(n) \leq f(n)$ for constants $c,k \geq 0$ and $n \geq k$.

\section{Background} \label{Section::Background}

The discrete time random walk can be described by the repeated
application of a unitary evolution operator $\uunpert$.  This
operator acts on a Hilbert space $\hs^{\coin}\tensorm\hs^{\shift}$
where $\hs^{\coin}$ is the Hilbert space associated with a quantum
coin and $\hs^{\shift}$ is the Hilbert space associated with the nodes
of the graph.  The operator $\uunpert$ can be written as
\cite{Aharonov:01}
\begin{equation}\label{Equation::WalkDef}
\uunpert = \shift\dotm\coin
\end{equation}
where $\shift$ is a permutation matrix which performs a controlled
shift based on the state of the quantum coin, and $\coin$ is a unitary
matrix which ``flips'' the quantum coin.  This operation can be
visualized by analogy to a classical random walk.  For each iteration
of a discrete time classical random walk on a graph, a coin is
flipped.  The walker then moves to an adjacent node specified by the
outcome of the coin flip.  An equivalent process occurs in the quantum
random walk, with the modification that the coin is a quantum coin and
can therefore exist in a superposition of states.  This modification
can lead to dramatic differences in behavior between the classical and
quantum random walks.  However, it should be noted that if the state
of the coin is measured after each flip, then the quantum random walk
reverts to a classical random walk (and similarly if the state of the
nodes is measured after every step).  

An important feature of the discrete time quantum random walk that has significance for its use in development of quantum algorithms, is that by virtue of its definition this walk will be efficiently implementable on a quantum computer whenever its classical counterpart is. (By efficient we mean that the walk can be simulated by a circuit with a number of gates that is polynomial in the number of bits (qubits)). This is due to the very similar structure of both these walks.  To illustrate this, assume we have an efficient way to implement the classical random walk on the underlying graph, i.e. to perform the coin-flip and subsequent shift. The shift is conditional on the outcome of the coin-flip (which determines the direction of the next step), i.e. we have a classical efficient circuit which performs a controlled shift on the basis states. It is straightforward \cite{Nielsen} to translate this circuit into a quantum circuit which performs the unitary controlled shift of \refeqn{WalkDef}. Similarly if there is an efficient procedure to flip the classical coin of the random walk, there will be an efficient way to implement a quantum coin. Hence implementation of the discrete time random walk is automatically efficient if the underlying classical walk is efficiently implementable. 

Note that if no measurement is
made, the quantum walk is controlled by a unitary operator rather than
a stochastic one.  This implies that there is no limiting stationary
distribution~\cite{Aharonov:01,Moore:01}. Nevertheless, several recent
works have shown that consistent notions of mixing time can be
formulated, and have shown polynomial speed up in these quantum mixing
times relative to the classical
analog~\cite{Aharonov:01,Moore:01}. Another quantity for which quantum
walks have shown speed-up relative to their classical analogs is the
hitting time~\cite{Kempe:02,Yamasaki:02}. Under certain conditions this
speed-up  can be exponential compared to the
classical analogue. We refer the reader to the recent papers
\cite{Aharonov:01}, \cite{Ambainis:01},
\cite{Moore:01} and \cite{Kempe:02} for some results
obtained from discrete time quantum random walks.

Our random walk search algorithm will be based on a random walk on the
$\numbit$-cube, i.e. the hypercube of dimension
$\numbit$ \cite{Kempe:02,Moore:01}.  The hypercube is a graph with
$N=2^\numbit$ nodes, each of which can be labelled by an $\numbit$-bit
binary string.  Two nodes on the hypercube described by bitstrings
$\xhc$ and $\vec{y}$ are are connected by an edge if
$\abs{\xhc-\vec{y}} = 1$, where $\abs{\xhc}$ is the Hamming weight of
$\xhc$.  In other words, if $\xhc$ and $\vec{y}$ differ by only a
single bit flip, then the two corresponding nodes on the graph are
connected.  Thus, each of the $2^\numbit$ nodes on the $\numbit$-cube
has degree $\numbit$ (i.e. it is connected to $\numbit$ other nodes),
so the Hilbert space of the algorithm is $\hs =
\hs^{\numbit}\tensorm\hs^{2^\numbit}$.  Each state in $\hs$ can be
described by a bit string $\xhc$, which specifies the position on the
hypercube, and a direction $\dhc$, which specifies the state of the
coin.  The shift operator, $\shift$, maps a state $\ket{\dhc, \xhc}$ onto the state $\ket{\dhc, \xhc \modp \vec{\dir_\dhc}}$, where
$\vec{\dir_\dhc}$ is the $\dhc^{th}$ basis vector on the hypercube.
$\shift$ can be written explicitly as,
\begin{equation}
\shift = \sum_{\dhc=0}^{\numbit-1}{\sum_{\xhc}{
\ket{\dhc,\xhc\modp \vec{\dir_{\dhc}}}\bra{\dhc,\xhc}}}
\end{equation}

To completely specify the unitary evolution operator $\uunpert$, the
coin operator, $\coin$, must also be chosen.  Normally, the coin
operator is chosen such that the same coin is applied to each node on
the graph.  This is the case in previous studies of discrete quantum
walks on the line~\cite{Watrous:01a,Aharonov:01,Ambainis:01} and on
the hypercube~\cite{Kempe:02,Moore:01}.  In other words, the coin
operator $\coin$ can be written as,
\begin{equation} \label{Equation::CoinUnperturbed}
\coin = \coin_0 \tensorm \id
\end{equation}
where $\coin_0$ is a $\numbit\by\numbit$ unitary operator acting on
$\hs^\coin$.  If $\coin$ is separable according to
\refeqn{CoinUnperturbed} then the eigenstates of $\uunpert$ are simply
the tensor product of the eigenstates of a (modified) coin
$\coin_{\vec{k}}$ and of the Fourier modes of the hypercube (labelled
by $\numbit$-bit strings $\vec{k}$)~\cite{Moore:01}. One frequently chosen
separable coin is Grover's ``diffusion'' operator on the coin space,
given by 
\begin{equation}
\coin_0  = \grover = -\id + 2\ket{\eqsupc}\bra{\eqsupc},
\end{equation}
where $\ket{\eqsupc}$ is the equal superposition over all $\numbit$
directions, i.e. $\ket{\eqsupc} =
\frac{1}{\sqrt{\numbit}}\sum_{\dhc=1}^\numbit{\ket{\dhc}}$
\cite{Moore:01}.  This coin operator is invariant to all permutations
of the $\numbit$ directions, so it preserves the permutation symmetry of
the hypercube.  The use of the Grover diffusion operator as a coin for
the hypercube was proposed in \cite{Moore:01}, where it was pointed
out that this operator is the permutation invariant operator farthest
away from the identity operator~\cite{Moore:01}. So, heuristically, it
should provide the most efficient mixing over states, from any given
initial state.  The non-trivial eigenvalues and eigenvectors of
$\uunpert$ are given by~\cite{Moore:01},
\begin{equation} \label{Equation::UnperturbedEV}
\e^{\pm i\eigunpert_{\kl}} = 1 - \frac{2\kl}{\numbit} \pm
\frac{2i}{\numbit}\sqrt{\kl\left(\numbit-\kl\right)}
\end{equation}
\begin{equation}
\ket{\eigunpertvec_{\khc}},\conj{\ket{\eigunpertvec_{\khc}}} =
\sum_{\xhc,\dhc}{\left(-1\right)^{\khc\dotm\xhc}\frac{2^{-
\numbit/2}}{\sqrt{2}}
\ket{\dhc,\xhc}\times
\cases{ 1/\sqrt{\kl} & if $\kl_\dhc = 1$ \cr
       \mp i/\sqrt{\numbit-\kl} & if $\kl_\dhc = 0$ \cr}}
\end{equation}
Note that the equal superposition over all states, $\ket{\inita} =
\ket{\eqsupc}\tensorm\ket{\eqsups}$, where $\ket{\eqsups}$ is the
equal superposition over the $2^\numbit$ nodes, is an eigenvector of
$\uunpert$ with eigenvalue $1$.  So repeated application of $\uunpert$
leaves the state $\ket{\inita}$ unchanged.

In order to create a search algorithm using the quantum random walk
architecture, we now consider a small perturbation of the unitary
operator $\uunpert$.  Specifically, we consider ``marking'' a single
arbitrary node by applying a special coin to that node.  In this
respect, the coin operator now takes on the function of an oracle.
Yet, contrary to the standard oracle which flips the phase of the marked state in the 
standard setting of a search algorithm, the oracle
acts instead by applying a marking coin, $\coin_1$, to the marked node
and a different coin, $\coin_0$, to the unmarked nodes.  Without loss
of generality, we can assume that the marked node corresponds to the all-zero string $\xhctarget = \vec{0}$.  Then our coin operator becomes
\begin{equation}
\coinpert = \coin_0\tensorm\id + \left(\coin_1 -
\coin_0\right)\tensorm\ket{\vec{0}}\bra{\vec{0}}.
\end{equation}
The marking coin, $\coin_1$, can be any $\numbit\by\numbit$
unitary matrix. For simplicity, we will consider here the case
where $\coin_1 = -\id$.
Then
our perturbed unitary evolution operator, $\upert$, is given by
\begin{equation}
\begin{array}{rcl}
\upert & = & \shift\dotm\coinpert\\
& = & \shift\dotm\left(\grover\tensorm\id -
\left(\grover-\id\right)\tensorm\ket{\vec{0}}\bra{\vec{0}}\right) \\
& = & \uunpert -
2\shift\dotm\left(\ket{\eqsupc}\bra{\eqsupc}\tensorm\ket{\vec{0}}\bra{\vec{0}}
\right)
\end{array}
\end{equation}
Analysis of the effects of this perturbation leads directly to the
definition of the random walk search algorithm, as will be
described
in the next section.

\section{Random Walk Search Algorithm} \label{Section::RWSA}

\subsection{Overview of the Algorithm}

We define the search space of the algorithm to be the set of all
$\numbit$-bit binary strings, $\xhc = \left\{0,1\right\}^\numbit$.  We
consider the function $\func\left(\xhc\right) = \left\{0,1\right\}$,
such that $\func\left(\xhc\right) = 1$ for exactly one input
$\xhctarget$.  Our goal is to find $\xhctarget$.  Using the mapping of
$\numbit$-bit binary string to nodes on the hypercube, this search
problem is then equivalent to searching for a single marked node amongst the $N=2^n$ nodes on
the $\numbit$-cube.  For purposes of the proof, we have set the marked
node to be $\xhctarget = \vec{0}$, but the location of the marked node has
no significance.

The random walk search algorithm is implemented as follows:
 \begin{enumerate}
\item Initialize the quantum computer to the equal superposition over
all states, $\ket{\inita} = \ket{\eqsupc}\tensorm\ket{\eqsups}$.  This can be 
accomplished efficiently on the node-space by applying $\numbit$ single-bit Hadamard operations to the $\ket{\vec{0}}$ state.  A similar procedure works for the direction space.
\item Given a coin oracle, $\coinpert$, which applies the coin
$\coin_0 = \grover$ to the unmarked states and the coin $\coin_1 =
-\id$ to the marked state, apply the perturbed evolution operator,
$\upert = \shift\dotm\coinpert$, $t_f=\frac{\pi}{2}\sqrt{2^\numbit}$ times.
\item Measure the state of the computer in the $\ket{\dhc,\xhc}$ basis.
\end{enumerate}

It is our claim that with probability $\frac{1}{2}-
\bigo\left(1/n\right)$, the outcome of the measurement will be the
marked state.  By repeating the algorithm a constant number of times,
we can determine the marked state with an arbitrarily small degree of
error.  In the remainder of this section we provide a proof of this
algorithm.

The general outline of the proof that we will present is the following. We need to determine the result of the operation $(\upert)^\time$ on the
initial state $\ket{\inita}$.  To do this, we will first simplify the
problem by showing that the perturbed walk on the hypercube can be
collapsed to a walk on the line (\refthm{Line}).  Next, by constructing two
approximate  eigenvectors of $\upert$, 
$\ket{\inita}$ and $\ket{\initb}$,
we will show that there are exactly two eigenvalues of
$\upert$ that are relevant (i.e. the initial state $\ket{\inita}$ has high overlap with the space spanned by the corresponding eigenvectors, see \refthm{SpectralGap} and \refthm{AtLeast}).
We denote these eigenvalues by $\e^{i\eigpert_0}$ and
$\e^{-i\eigpert_0}$.  We will then show that the corresponding
eigenvectors $\ket{\eigpert_0}$ and $\ket{- \eigpert_0}$ can be
well-approximated by linear combinations of the initial state $\ket{\inita}$ and the 
second state $\ket{\initb}$ (\refthm{InitaElements}).  As a result, our random walk search
algorithm can be approximated by a two-dimensional rotation in the
$\ket{\eigpert_0},\ket{-\eigpert_0}$-plane away from the initial state
$\ket{\inita}\approx 1/\sqrt{2}(\ket{\eigpert_0}+\ket{-\eigpert_0})$ and towards $\ket{\initb} \approx i/\sqrt{2}(- \ket{\eigpert_0}+\ket{-\eigpert_0})$, which constitutes a very close approximation to the
target state $\ket{\xhctarget}$. Finally, we show that each
application of the evolution operator $\upert$ corresponds to a
rotation angle of approximately $1/\sqrt{2^{\numbit-1}}$ (\refthm{EigImpGap}).  Hence, the
search is completed after approximately
$\frac{\pi}{2}\sqrt{2^{\numbit-1}}$ steps, {\it i.e.}, after
$\bigo\left(\sqrt{\numdim}\right)$ calls to the oracle, where $N=2^n$
is the number of nodes.
\subsection{Proof of Correctness}
In general, analytic determination of the eigenspectrum of a large
matrix is a daunting task, so we will take advantage of the symmetries
inherent in $\upert$ to simplify the problem. Let us first show that the
perturbed random walk on the hypercube can be collapsed onto a random
walk on the line. Let $\swapij$ be the permutation operator which
swaps the bits $i$ and $j$, in both the node space and the coin space.
In other words, given a state $\ket{\dhc,\xhc}$, under the permutation
operator, $\swapij$, the $i^{th}$ and $j^{th}$ bits of $\xhc$ are
swapped and the directions $\dhc=i$ and $\dhc=j$ are swapped.  Clearly
the unperturbed evolution operator $\uunpert$ commutes with $\swapij$
since every direction in the unperturbed walk is equivalent.

\begin{theorem}\label{Theorem::Line}
$\upert$ commutes with $\swapij$.
\end{theorem}

\emph{Proof.}  \begin{equation} \begin{array}{rcl}
\swapij^\dagger\upert\swapij &
= & \swapij^\dagger\uunpert\swapij -
2\swapij^\dagger\shift\dotm\left(\ket{\eqsupc}\bra{\eqsupc}\otimes \ket{\vec{0}}\bra{\vec{0}}\right)\swapij\\
& = & \uunpert -
\frac{2}{\sqrt{\numbit}}\sum_{\dhc=0}^{\numbit-
1}{\swapij^\dagger\ket{\dhc,\vec{\dir_\dhc}}\bra{\dhc,0}}\swapij
\\ & = & \upert 
\end{array} 
\end{equation} 
So,
$\commutator{\upert}{\swapij} = 0$. \bbox

Because the initial state $\ket{\inita}$ is an eigenvector of
$\swapij$ with eigenvalue $1$ for all $i$ and $j$, and
$\commutator{\upert}{\swapij} = 0$, any intermediate state $\ket{\psi_\time} =
(\upert)^{\time}\ket{\inita}$ must also be an eigenvector of eigenvalue
$1$ with respect to $\swapij$.  Thus, $(\upert)^\time$ preserves the
symmetry of $\ket{\inita}$ with respect to bit swaps.  It is therefore
useful to define $2\numbit$ basis states,
$\ket{R,0},\ket{L,1},\ket{R,1},\ldots,\ket{R,n-1},\ket{L,n}$ where
\begin{eqnarray}
\ket{R,\xl} & = &
\sqrt{\frac{1}{\left(\numbit-
\xl\right)\nchoosek{n}{x}}}\sum_{\dist{\xhc} =
\xl}{\sum_{\xl_\dhc = 0}{\ket{\dhc,\xhc}}} \\
\ket{L,\xl} & = &
\sqrt{\frac{1}{\xl\nchoosek{n}{x}}}\sum_{\dist{\xhc} =
\xl}{\sum_{\xl_\dhc = 1}{\ket{\dhc,\xhc}}}
\end{eqnarray}
which are also invariant to bit swaps $\swapij$.  These states span the 
eigenspace of eigenvalue $1$ of $\swapij$. Using these basis states, we
can project out all but one spatial degree of freedom and effectively
reduce the random walk on the hypercube to a random walk on the line.
This is illustrated in \reffig{CubeToLine}.  The marked node
corresponds now to $\ket{R,0}$.  We can rewrite $\uunpert$, $\upert$,
and $\ket{\inita}$ in this collapsed basis. First note that the shift operator $\shift$ in this basis acts as
\begin{equation}
\shift = \sum_{x=0}^{n-1} \ket{R,x} \bra{L,x+1}+\ket{L,x+1} \bra{R,x} 
\end{equation} 
and the unperturbed coin acts as 
\begin{equation} 
\coin_0 = \sum_{x=0}^n \left(\begin{array}{cc} \cos \omega_x & \sin \omega_x \\ \sin \omega_x & -\cos \omega_x \end{array} \right)  \otimes \ket{x} \bra{x} 
\end{equation}
where $\cos \omega_{\xl} = 1 - \frac{2\xl}{\numbit}$
and $\sin \omega_{\xl} = \frac{2}{\numbit}\sqrt{\xl\left(\numbit-\xl\right)}$ and where the first part acts on the space spanned by $\{\ket{R},\ket{L}\}$ and the second part acts on the positions $\{\ket{0},\ldots,\ket{n}\}$ on the line. Note that the coin of the collapsed walk is not homogeneous in space any more. The unitary operator $U$ on the restricted space acts as
\begin{equation} \label{Equation::CollapsedU}
\begin{array}{rcl}
\uunpert &=& \sum_{\xl=0}^{\numbit-1}{
\ket{R,\xl}\left(-\cos \omega_{\xl+1} \bra{L,\xl+1} +
\sin \omega_{\xl+1} \bra{R,\xl+1}\right)} + \\
& & \sum_{\xl=1}^{\numbit}{
\ket{L,\xl}\left(\sin \omega_{\xl-1} \bra{L,\xl-1} +
\cos \omega_{\xl-1} \bra{R,\xl-1}\right)}
\end{array}.
\end{equation}
Similarly,
\begin{equation} \label{Equation::UpertCollapsed}
\upert = \uunpert + \Delta\uunpert = \uunpert -
2\ket{L,1}\bra{R,0}
\end{equation}
Note that the only difference between $\uunpert$ and $\upert$ is in the
sign of the matrix element in position ($\ket{L,1},\ket{R,0}$).  Finally,
\begin{equation} \label{Equation::InitaCollapsed}
\ket{\inita} = \frac{1}{\sqrt{2^\numbit}}\ket{R,0} +
\frac{1}{\sqrt{2^\numbit}}\ket{L,\numbit} +
\sum_{\xl=1}^{\numbit-1}{\left(
\sqrt{\frac{\nchoosek{\numbit-1}{\xl-1}}{2^\numbit}}\ket{L,\xl}+
\sqrt{\frac{\nchoosek{\numbit-
1}{\xl}}{2^\numbit}}\ket{R,\xl}\right)}
\end{equation}
Since $\uunpert$ and $\swapij$ are mutually diagonalizable, the eigenvectors 
of 
$\uunpert$ in the reduced space are also bit-flip invariant.  
Examining \refeqn{UnperturbedEV}, it is clear that if we take the equal
superpositions of all  eigenvectors of same eigenvalue
$\ket{\eigunpertvec_{\khc}}$ such that $\abs{\khc} = \kl$, the
resulting eigenvector will be bit-swap invariant.  Thus we define,
\begin{equation} \label{Equation::UnperturbedEVSym}
\ket{\eigunpert_{\kl}} =
\frac{1}{\sqrt{\nchoosek{\numbit}{\kl}}}\sum_{\abs{\khc}=\kl}{\ket
{\eigunpertvec_{\khc}}}
\end{equation}
which are the eigenvectors of $\uunpert$ with eigenvalues
$\e^{i\eigunpert_{\kl}}$ in the collapsed (symmetric) space.

Note that both $\uunpert$ and $\upert$ are represented by real
matrices; therefore, their eigenvalues and eigenvectors will come in
complex conjugate pairs.

Having determined these general properties of the perturbed matrix
$\upert$, we now turn to the problem of analyzing the eigenvalue
spectrum of $\upert$.
Let $\arcimp$ be the arc on the unit circle containing all complex
numbers of unit norm with real part greater than $1-2/3\numbit$.  In
other words,
\begin{equation}
\arcimp = \left\{\,z \mid \real z > 1-\frac{2}{3\numbit}, \abs{z} =
1\right\}
\end{equation}
\reffig{NumericalSpectra} shows the geometrical representation of
$\arcimp$ together with the eigenvalue spectra of the unperturbed and
perturbed matrices for $\numbit = 8$.  We will prove that $\arcimp$
contains exactly two eigenvalues $\e^{i\eigpert_0}$ and
$\e^{-i\eigpert_0}$ of $\upert$.  First, we will prove that there are
\emph{at most} two eigenvalues with real part greater than
$1-2/3\numbit$.  Then we will show that there are \emph{at least} two eigenvalues on 
$\arcimp$.  From these facts, it follows that there are exactly two eigenvalues of 
$\upert$ on $\arcimp$.

\begin{theorem} \label{Theorem::SpectralGap}
There are at most two eigenvalues of $\upert$ with real part greater
than $1- 2/3\numbit$.
\end{theorem}

\emph{Proof.}  We will prove by contradiction.  Let us assume that
there are three eigenvalues, $\e^{i\eigpert_0}$,$\e^{i\eigpert_1}$,
and $\e^{i\eigpert_2}$, with real part greater than $1-2/3\numbit$.
Let $\ket{\eigpert_0}$, $\ket{\eigpert_1}$, and $\ket{\eigpert_2}$ be
the corresponding eigenvectors.  Then,
\begin{equation} \label{Equation::ReSumEigpert}
\begin{array}{rcl}
\real \sum_i{\bra{\eigpert_i}\upert\ket{\eigpert_i}} &=&
\real \sum_i{\e^{i\eigpert_i}\bkm{\eigpert_i}{\eigpert_i}} \\
&>& 3-2/\numbit
\end{array}
\end{equation}
Let us define $\subspaceimp$ to be the subspace spanned by
$\ket{\eigpert_0}$, $\ket{\eigpert_1}$, $\ket{\eigpert_2}$.  Then we
can write \refeqn{ReSumEigpert} as the partial trace of $\upert$ over
$\subspaceimp$,
\begin{equation} \label{Equation::ReTrUpert}
\real \trace_{\subspaceimp}\upert > 3-2/\numbit
\end{equation}

Let us now define $\ket{\initm} = \frac{1}{\sqrt{2}}\left(\ket{0,R} -
\ket{1,L}\right)$.  We can expand the $\ket{\eigpert_0}$,
$\ket{\eigpert_1}$, and $\ket{\eigpert_2}$ in terms of $\ket{\inita}$,
$\ket{\initm}$ and a residual vector,
\begin{eqnarray}
\ket{\eigpert_0} &=& \coeffb_{00}\ket{\inita} +
\coeffb_{01}\ket{\initm} +
\coeffb_{02}\ket{\resb_0} \nonumber \\
\ket{\eigpert_1} &=& \coeffb_{10}\ket{\inita} +
\coeffb_{11}\ket{\initm} + \coeffb_{11}\ket{\resb_1} \nonumber \\
\ket{\eigpert_2} &=&
\coeffb_{20}\ket{\inita} + \coeffb_{21}\ket{\initm} +
\coeffb_{22}\ket{\resb_2} \end{eqnarray}
where $\ket{\resb_i}$ is a normalized vector orthogonal to
$\ket{\inita}$ and $\ket{\initm}$.  We now observe that, due to the
basis invariance of the trace, \refeqn{ReTrUpert} holds for any linear
combination of $\ket{\eigpert_0}$, $\ket{\eigpert_1}$, and
$\ket{\eigpert_2}$.  Thus, we can construct three new orthonormal vectors,
$\ket{\eigtrial_0}$, $\ket{\eigtrial_1}$, and $\ket{\eigtrial_2}$ by
taking linear combinations of $\ket{\eigpert_0}$, $\ket{\eigpert_1}$,
and $\ket{\eigpert_2}$ such that,
\begin{equation} \label{Equation::EigtrialConstraints}
\bkm{\eigtrial_2}{\inita} = \bkm{\eigtrial_2}{\initm} = 0
\end{equation}
In other words, we can expand $\ket{\eigtrial_0}$,
$\ket{\eigtrial_1}$, and $\ket{\eigtrial_2}$ as,
\begin{equation} \label{Equation::EigtrialExpand}
\begin{array}{r c c c c}
\ket{\eigtrial_0} &=& \coeff_{00}\ket{\inita}&
+\coeff_{01}\ket{\initm} &
+\coeff_{02}\ket{\res_0} \\
\ket{\eigtrial_1} &=& \coeff_{10}\ket{\inita}&
+\coeff_{11}\ket{\initm}&
+\coeff_{12}\ket{\res_1}
\\
\ket{\eigtrial_2} &=& & & \quad\quad\, \ket{\res_2}
\end{array}
\end{equation}
Since $\ket{\eigtrial_0}$, $\ket{\eigtrial_1}$, and
$\ket{\eigtrial_2}$ still form a basis for $\subspaceimp$, from 
\refeqn{ReTrUpert} 
it follows that,
\begin{equation} \label{Equation::AssumptionInequality}
\begin{array}{rcl}
3-2/\numbit &<&
\real \sum\limits_i{\bra{\eigtrial_i}\upert\ket{\eigtrial_i}} 
\end{array}
\end{equation}
Since $\upert$ is a unitary operator, we know that
$\real\bra{\eigtrial_i}\upert\ket{\eigtrial_i} \leq 1$ for all
$\ket{\eigtrial_i}$.  Thus, applying this inequality to the first two
terms in the sum, we obtain,
\begin{equation} \label{Equation::MaxUpertTwoEigtrials}
\real \sum_i{\bra{\eigtrial_i}\upert\ket{\eigtrial_i}} \leq 2 +
\real \bra{\eigtrial_2}\upert\ket{\eigtrial_2}
\end{equation}
Since, $\upert = \uunpert + \Delta\uunpert$, we can write
\begin{equation}
\real\bra{\eigtrial_2}\upert\ket{\eigtrial_2} =
\real\bra{\eigtrial_2}\uunpert\ket{\eigtrial_2} +
\real\bra{\eigtrial_2}\Delta\uunpert\ket{\eigtrial_2}
\end{equation}
Let us first consider $\bra{\eigtrial_2}\uunpert\ket{\eigtrial_2}$.
We can expand $\ket{\eigtrial_2}$ in terms of the unperturbed eigenstates,
$\ket{\eigtrial_2} = \sum_j{\upcoeff_j\ket{\eigunpert_j}}.$
So,
$\real\bra{\eigtrial_2}\uunpert\ket{\eigtrial_2} =
\sum_j{\abs{\upcoeff_j}^2\cos{\eigunpert_j}}.$
However, since $\bkm{\eigtrial_2}{\inita} = 0$, there is no
contribution from the eigenvalue with value $1$.  The eigenvalue with
the next- largest real part is $\e^{i\eigunpert_1} = 1 - 2/\numbit + i
\frac{2}{\numbit}{\sqrt{\numbit-1}}$.  Thus,
\begin{equation} \label{Equation::MaxUunpertEigtrial}
\real\bra{\eigtrial_2}\uunpert\ket{\eigtrial_2}
\leq
1-2/\numbit
\end{equation}
Next, we consider, $\bra{\eigtrial_2}\Delta\uunpert\ket{\eigtrial_2}$.
Let $\ket{\initp} = \frac{1}{\sqrt{2}}\left(\ket{0,R} +
\ket{1,L}\right)$.  Using \refeqn{UpertCollapsed} we can express
$\Delta\uunpert$ in terms of $\ket{\initm}$ and $\ket{\initp}$,
\begin{equation}
\Delta\uunpert =
\ket{\initm}\bra{\initm} + \ket{\initm}\bra{\initp} -
\ket{\initp}\bra{\initm} - \ket{\initp}\bra{\initp}
\end{equation}
But since $\bkm{\eigtrial_2}{\initm} = 0$ (see
\refeqn{EigtrialConstraints}),
\begin{equation} \label{Equation::MaxDeltaUunpertEigtrial}
\bra{\eigtrial_2}\Delta\uunpert\ket{\eigtrial_2} =
\left(-
\abs{\bkm{\initp}{\eigtrial_2}}^2\right)
\leq 
0
\end{equation}
Then,
$\real \bra{\eigtrial_2}\upert\ket{\eigtrial_2}
\leq 1-2/\numbit.$
Combining \refeqn{MaxUpertTwoEigtrials}, \refeqn{MaxUunpertEigtrial},
and \refeqn{MaxDeltaUunpertEigtrial}, we obtain,
\begin{equation}
\real \sum_i{\bra{\eigtrial_i}\upert\ket{\eigtrial_i}} \leq 3-2/\numbit\ .
\end{equation}
Since this contradicts \refeqn{AssumptionInequality}, our assumption
must be false.  \bbox

\begin{theorem}\label{Theorem::AtLeast}
There are at least two eigenvalues of $\upert$ on $\arcimp$.
\end{theorem}
\emph{Proof.}
We will construct two approximate eigenvectors of $\upert$, $\ket{\inita}$ and
$\ket{\initb}$.  $\ket{\inita}$ is given by \refeqn{InitaCollapsed}.
Using \refeqn{UpertCollapsed},
\begin{equation} \label{Equation::UpertInita}
\upert\ket{\inita} = \ket{\inita} - 2/\sqrt{2^{\numbit}}\ket{L,1}
\end{equation}
And,
\begin{equation} \label{Equation::InitaUpertInita}
\begin{array}{rcl}
\bra{\inita}\upert\ket{\inita} &=& \bkm{\inita}{\inita} -
\bkm{\inita}{L,1}\bkm{R,0}{\inita} \\
&=& 1-1/2^{\numbit-1}
\end{array}
\end{equation}
So, apart from a small residual, $\ket{\inita}$ is also ``almost'' an
eigenvector of $\upert$ with eigenvalue $1$.  Now, we need to find a 
second approximate eigenvector, $\ket{\initb}$.  Let
\begin{equation}\label{Equation::InitbCollapsed}
\ket{\initb} =
\left(\sum_{\xl=0}^{\numbit/2-1}{
\frac{1}{\sqrt{2\nchoosek{\numbit-1}{\xl}}}\ket{R,\xl} -
\frac{1}{\sqrt{2\nchoosek{\numbit-1}{\xl}}}\ket{L,\xl+1}
}\right)/\norminitb
\end{equation}
where $\norminitb$ is a normalization constant,
\begin{equation}
\norminitb =
\sqrt{\sum_{\xl=0}^{\numbit/2-1}{\frac{1}{\nchoosek{\numbit-
1}{\xl}}}}
\end{equation}
Using this definition and \refeqn{CollapsedU}, we see that
\begin{equation} \label{Equation::UpertInitb}
\upert\ket{\initb} = \ket{\initb} -
\frac{1}{\norminitb\sqrt{2\nchoosek{\numbit-
1}{\numbit/2}}}\left(\ket{R,\numbit/2-1}
+ \ket{L,\numbit/2+1}\right)
\end{equation}
So,
\begin{equation} \label{Equation::InitbUpertInitb}
\bra{\initb}\upert\ket{\initb} = 1 -
\frac{1}{2\norminitb^2\nchoosek{\numbit-1}{\numbit/2}}
\end{equation}
It is straightforward to show that $1 < \norminitb^2 < 1+2/\numbit$ for sufficiently large $n$.
Thus, except for a small residual, $\ket{\initb}$ is ``almost'' an
eigenvector of $\upert$ with eigenvalue $1$.

Now let us verify that there is at least one
eigenvalue of $\upert$ on $\arcimp$.  Let us assume that there are no
eigenvalues of $\upert$ on $\arcimp$.  Then $\cos{\eigpert_j} <
1-2/3\numbit$ for all $j$.  Then using \refeqn{InitaUpertInita},
\begin{equation}
\begin{array}{rcl}
1-1/2^{\numbit-1} &=& \real\bra{\inita}\upert\ket{\inita} \\
&=&
\sum\limits_j{\abs{\bkm{\inita}{\eigpert_j}}^2\cos{\eigpert_j}} \\
&<&
\left(1-
2/3\numbit\right)\sum\limits_j{\abs{\bkm{\inita}{\eigpert_j}}^2} \\
&=& 1-2/3\numbit
\end{array}
\end{equation}
which is wrong for $\numbit >3$. Hence our
assumption is false and there must be at least one eigenvalue of
$\upert$ on $\arcimp$.

Now let us assume there is exactly one eigenvalue of
$\upert$, $\e^{i\eigpert_0}$, on $\arcimp$.  Then,
\begin{equation}
\begin{array}{rcl}
1-\frac{1}{2^{\numbit-1}} &=&
\real\bra{\inita}\upert\ket{\inita} \\
&=& \sum_j{\abs{\bkm{\inita}{\eigpert_j}}^2 \cos{\eigpert_j}}\\
&=& \abs{\bkm{\inita}{\eigpert_0}}^2\cos{\eigpert_0} + \sum_{j
\neq 0}
{\abs{\bkm{\inita}{\eigpert_j}}^2 \cos{\eigpert_j}}\\
&\leq&
\abs{\bkm{\inita}{\eigpert_0}}^2 +
\left(1-\abs{\bkm{\inita}{\eigpert_0}}^2\right)\left(1-
2/3\numbit\right)\\
\end{array}
\end{equation}
Rearranging terms,
\begin{equation}
\abs{\bkm{\inita}{\eigpert_0}}^2 \geq 1-
\frac{3\numbit}{2^{\numbit}}
\end{equation}
If we use $\ket{\initb}$ as a trial vector and follow the same
arguments, we obtain the inequality
\begin{equation}
\abs{\bkm{\initb}{\eigpert_0}}^2 \geq
1-\frac{3\numbit}{4\norminitb^2\nchoosek{\numbit-1}{\numbit/2}}.
\end{equation}
But since $\ket{\inita}$ and $\ket{\initb}$ are orthonormal, this leads
to a contradiction, since,
\begin{equation}
\begin{array}{rcl}
1 &=& \bkm{\eigpert_0}{\eigpert_0}\\
&\geq& \abs{\bkm{\inita}{\eigpert_0}}^2 +
\abs{\bkm{\initb}{\eigpert_0}}^2 \\
&\geq& 2 - \frac{3\numbit}{2^{\numbit}} -
\frac{3\numbit}{4\norminitb^2\nchoosek{\numbit-1}{\numbit/2}}
\end{array}
\end{equation}
which is not true for large $\numbit$.  Hence, there must
be at least two eigenvalues on the arc $\arcimp$.  \bbox

As noted above, the eigenvalues and eigenvectors of $\upert$ come in
complex conjugate pairs. In particular the two
eigenvalues on $\arcimp$ must be a complex conjugate pair;  let
$\e^{\pm i\eigpert_0}$ be the two eigenvalues on $\arcimp$.  The corresponding eigenvectors obey $\ket{-\eigpert_0} = \conj{\ket{\eigpert_0}}$ (if $\e^{i\eigpert_0}=\e^{- i\eigpert_0}=1$,
then we can construct linear combinations of $\ket{\eigpert_0}$ and
$\ket{-\eigpert_0}$ for which this statement is true). We will
now show that $\ket{\pm \eigpert_0}$ can be well-approximated by
linear combinations of $\ket{\inita}$ and $\ket{\initb}$.

\begin{theorem} \label{Theorem::InitaElements}
\begin{eqnarray} \label{Equation::KetEigimp}
\ket{\eigpert_0} &=& \sqrt{\prob_0}\ket{\inita} +
\sqrt{\prob_1}\e^{i\abphase}\ket{\initb}
+ \sqrt{1-\prob_0-\prob_1}
\ket{\res_0}\nonumber\\
\ket{-\eigpert_0} &=& \sqrt{\prob_0}\ket{\inita} +
\sqrt{\prob_1}\e^{-i\abphase}\ket{\initb}
+ \sqrt{1-\prob_0-\prob_1}
\conj{\ket{\res_0}}
\end{eqnarray}
where $\prob_0 = \abs{\bkm{\eigpert_0}{\inita}}^2 = \abs{\bkm{-
\eigpert_0}{\inita}}^2$, $\prob_1 = \abs{\bkm{\eigimp_0}{\initb}}^2 = \abs{\bkm{-\eigimp_0}{\initb}}^2$ and $\ket{\res_0}$ is a normalized vector orthogonal to $\ket{\inita}$ and $\ket{\initb}$. Furthermore, 
$1/2 \geq \prob_0 \geq 1/2 - 3n/2^{\numbit+1}$ and $1/2 \geq \prob_1
\geq 1/2 - \frac{3n}{8 c^2 \nchoosek{\numbit-1}{\numbit/2}}$, with $\e^{i\abphase} = i + \Delta$, where $\abs{\Delta}=\bigo\left(\frac{\numbit}{\nchoosek{\numbit-1}{\numbit/2}}\right)$.
\end{theorem}

\emph{Proof.}
Since
$\ket{\inita}$ and $\ket{\initb}$ are real vectors, $\abs{\bkm{\eigpert_0}{\inita}}^2 =
\abs{\bkm{-\eigpert_0}{\inita}}^2\leq 1/2$ and $\abs{\bkm{\eigpert_0}{\initb}}^2 =
\abs{\bkm{-\eigpert_0}{\initb}}^2\leq 1/2$.
Using \refeqn{InitaUpertInita},
\begin{equation}
\begin{array}{rcl}
1-\frac{1}{2^{\numbit-1}} &=& \real
\bra{\inita}\upert\ket{\inita}\\
&=&\sum_j{\cos \eigpert_j}\abs{\bkm{\eigpert_j}{\inita}^2}\\
&=&
2\prob_0\cos{\eigpert_0}
+
\sum_{j\neq0}{\cos{\eigpert_j}\abs{\bkm{\eigpert_j}{\inita}}^2}\\
&<& 2\prob_0 + \left(1-2/3\numbit\right)\left(1-2\prob_0\right)
\end{array}
\end{equation}
Rearranging terms, we obtain,
\begin{equation}
\prob_0 \geq 1/2 - \frac{3\numbit}{2^{\numbit+1}}.
\end{equation}
Using \refeqn{InitbUpertInitb} and the same arguments as above we obtain 
\begin{equation}
\prob_1 \geq 1/2 -
\frac{3\numbit}{8\norminitb^2\nchoosek{\numbit-1}{\numbit/2}}.
\end{equation}
Up to a global phase $\ket{\eigpert_0}$ can be written as,
\begin{equation}
\ket{\eigpert_0} = \abs{\bkm{\eigpert_0}{\inita}} \ket{\inita} +
\abs{\bkm{\eigpert_0}{\initb}}\e^{i\abphase}\ket{\initb}
+ \sqrt{1-\prob_0-\prob_1}
\ket{\res_0}
\end{equation}
which yields \refeqn{KetEigimp} for $\ket{\eigpert_0}$ and $\ket{-\eigpert_0}$.

To estimate $e^{i \eta}$ note that since $\ket{\eigpert_0}$ and $\ket{-\eigpert_0}$ are
eigenvectors of a unitary matrix, they must be orthogonal.
Consequently,
\begin{equation} \label{Equation::ABPhase}
\begin{array}{rcl}
0 &=& \bkm{-\eigpert_0}{\eigpert_0} \\
&=& \prob_0 + \prob_1 \left(\e^{i\abphase}\right)^2 +(1-\prob_0 -\prob_1) \bkm{\conj{\res_0}}{\res_0}
\end{array}
\end{equation}
Solving for $\e^{i\abphase}$, we obtain,
\begin{eqnarray}
\real \left(\e^{i\abphase}\right)^2 &=& \frac{-\prob_0 - (1-\prob_0-\prob_1)\real \bkm{\conj{\res_0}}{\res_0}}{\prob_1}
\end{eqnarray}
Assume $\real \bkm{\conj{\res_0}}{\res_0} \geq 0$. Then, using $1/(1-x) \leq 1+2x$ for small $x$ we get
\begin{eqnarray}
-1+\frac{3n}{2^n}&\geq& -\frac{\prob_0}{\prob_1} \nonumber\\
&\geq&\real  \left(\e^{i\abphase}\right)^2 \geq 
-\frac{\prob_0+\left(\frac{3\numbit}{2^{\numbit+1}} + 
\frac{3\numbit}{8\norminitb^2\nchoosek{\numbit-1}{\numbit/2}}\right)}{\prob_1}\nonumber\\
&\geq& -1-2\left(\frac{3\numbit}{2^{\numbit+1}} +
\frac{3\numbit}{8\norminitb^2\nchoosek{\numbit-1}{\numbit/2}}\right)-4 \left(\frac{3\numbit}{8\norminitb^2\nchoosek{\numbit-
1}{\numbit/2}}\right)
\end{eqnarray}
which in turn implies that $e^{i \eta}=i+\Delta$ with $\abs{\Delta}=\bigo\left(\frac{\numbit}{\nchoosek{\numbit-1}{\numbit/2}}\right)$. A similar reasoning holds if $\real \bkm{\conj{\res_0}}{\res_0} \leq 0$. \bbox

As a last ingredient we need to bound the angle $\eigpert_0$. These bounds are provided in the following final theorem.
\begin{theorem} \label{Theorem::EigImpGap}
$-\frac{1}{\norminitb\sqrt{2^{\numbit-1}}} - \eigimpres
\leq
\eigpert_0
\leq -\frac{1}{\norminitb\sqrt{2^{\numbit-1}}} + \eigimpres$,
where
$\eigimpres =
\bigo\left(\frac{\numbit^{3/2}}{2^\numbit}\right)$
\end{theorem}

\emph{Proof.}  
We will approximate $\e^{i\eigpert_0}=\bra{\eigpert_0}\upert \ket{\eigpert_0}$ by $\bra{\eigtrial}\upert\ket{\eigtrial}$, where $\ket{\eigtrial} =
1/\sqrt{2}\left(\ket{\inita} + \e^{i\abphase}\ket{\initb}\right)$. Let us first evaluate $\abs{\e^{i\eigpert_0} -
\bra{\eigtrial}\upert\ket{\eigtrial}}$.  We
can expand $\upert$ in terms of its eigenvectors to obtain,
\begin{equation}
\abs{\e^{i\eigpert_0} - \bra{\eigtrial}\upert\ket{\eigtrial}} =
\abs{\e^{i\eigpert_0} -
\sum_j{\abs{\bkm{\eigpert_j}{\eigtrial}}^2\e^{i\eigpert_j}}}
\end{equation}
We then note that from \refeqn{KetEigimp},
\begin{equation}
\abs{\bkm{\eigpert_0}{\eigtrial}}^2 = \sqrt{\prob_0/2} +
\sqrt{\prob_1/2}
\end{equation}
So,
\begin{equation}\label{Equation::EigImpGap}
\begin{array}{rcl}
\abs{\e^{i\eigpert_0} - \bra{\eigtrial}\upert\ket{\eigtrial}} &=&
\abs{\e^{i\eigpert_0} -
\left(\sqrt{\prob_0/2} + \sqrt{\prob_1/2}\right)\e^{i\eigpert_0}
+
\sum\limits_{\ket{\eigpert_j}\neq\ket{\eigpert_0}}{\abs{\bkm{\eigpert_j}{\eigtrial}}^2\e^{i\eigpert_j}}}\\
&\leq&
\abs{\e^{i\eigpert_0}
\left(1-\sqrt{\prob_0/2} - \sqrt{\prob_1/2}\right)}
+
\sum\limits_{\ket{\eigpert_j}\neq\ket{\eigpert_0}}{\abs{\bkm{\eigpert_j}{\eigtrial}}^2}\\
&\leq&
2\left(1-\sqrt{\prob_0/2} - \sqrt{\prob_1/2}\right)\\
&\leq&
2\left(\frac{3\numbit}{2^{\numbit+1}} +
\frac{3\numbit}{8\norminitb^2\nchoosek{\numbit-1}{\numbit/2}}\right)
\end{array}
\end{equation}
with 
$\sqrt{1-x}\geq 1-x$ for $0\leq x \leq 1$.
Using the fact that the binomial coefficients approach the Gaussian
distribution for large $\numbit$, such that,
\begin{equation} \label{Equation::LimitBinomial}
\nchoosek{\numbit}{\xl} =
\sqrt{\frac{2}{\pi\numbit}}\e^{-\frac{x-
\numbit/2}{\numbit/2}}2^\numbit
\end{equation}
we can rewrite \refeqn{EigImpGap} taking the leading order terms in
$\numbit$.  Recalling that $\norminitb > 1$, we obtain,
\begin{equation} \label{Equation::EigImpGapB}
\abs{\e^{i\eigpert_0} - \bra{\eigtrial}\upert\ket{\eigtrial}} =
\bigo \left(\frac{\numbit^{3/2}}{2^\numbit}\right)
\end{equation}
\refeqn{EigImpGapB} is an explicit formula which bounds the distance in the complex 
plane between the eigenvalue of interest, $\e^{i\eigpert_0}$, and the matrix element 
$\bra{\eigtrial}\upert\ket{\eigtrial}$.  \reffig{EigImpGap} shows the 
geometric 
representation of \refeqn{EigImpGapB}.  
Note that,
\begin{equation}
\abs{\sin{\eigpert_0} - \imag \bra{\eigtrial}\upert\ket{\eigtrial}} = \abs {\imag \left(e^{i{\eigpert_0}} - \bra{\eigtrial}\upert\ket{\eigtrial}\right)} \leq 
\abs{e^{i{\eigpert_0}} - \bra{\eigtrial}\upert\ket{\eigtrial}}
\end{equation}
Next, we evaluate 
$\imag \bra{\eigtrial}\upert\ket{\eigtrial}$ using
\refeqs{UpertInita}{UpertInitb},
\begin{equation} \label{Equation::EigtrialUpertEigtrial}
\begin{array}{rcl}
\imag \bra{\eigtrial}\upert\ket{\eigtrial} &=&
\imag \left(\e^{i\abphase}\bra{\inita}\upert\ket{\initb}-
\e^{i\abphase}\bra{\initb}\upert\ket{\inita}\right)\\
&=&
\imag \frac{1}{2}\left(
-\frac{\e^{i\abphase}}{\norminitb\sqrt{2\nchoosek{\numbit-
1}{\numbit/2}}}\left(\bkm{\inita}{R,\numbit/2-
1}+\bkm{\inita}{L,\numbit/2+1}\right)
-\e^{i\abphase}\left(-
\frac{2}{\sqrt{2^\numbit}}\bkm{\initb}{L,1}\right)\right)\\
&=&
-\imag \frac{\e^{i\abphase}}{\norminitb\sqrt{2^{\numbit-1}}}=-\frac{1}{\norminitb\sqrt{2^{\numbit-1}}} -\bigo\left(\frac{\numbit}{\sqrt{2^\numbit}\nchoosek{\numbit-1}{\numbit/2}}\right) \\
\end{array}
\end{equation}
Then, using \refthm{InitaElements}, \refeqn{EigImpGapB}, and \refeqn{EigtrialUpertEigtrial} we 
can write
\begin{equation}
\abs{\sin{\eigpert_0}
+ \frac{1}{\norminitb\sqrt{2^{\numbit-1}}}
+ \bigo\left(\frac{\numbit}{\sqrt{2^\numbit}\nchoosek{\numbit-1}{\numbit/2}}\right)}
= \bigo\left(\frac{\numbit^{3/2}}{2^\numbit}\right)
\end{equation}
Using $\sin{x} = x + \bigo (x^3)$ and keeping only 
leading order terms solving for $\eigpert_0$ gives us,
\begin{equation} \label{Equation::EigImpAngle}
-\frac{1}{\norminitb\sqrt{2^{\numbit-1}}}
-\bigo\left(\frac{\numbit^{3/2}}{2^\numbit}\right)
\leq
 \eigpert_0 \leq
-\frac{1}{\norminitb\sqrt{2^{\numbit-1}}}+
\bigo\left(\frac{\numbit^{3/2}}{2^\numbit}\right).\quad \bbox
\end{equation}

We can now quantitatively describe the overall operation of the algorithm.
Starting with initial state $\ket{\inita}$, we consider the state of
the computer after $\time$ applications of $\upert$.  Then using
\refthm{InitaElements} we can expand $\ket{\inita}$ as
\begin{equation}
\ket{\inita} = \sqrt{\prob_0} (\ket{\eigpert_0}+\ket{-\eigpert_0})+\delta \ket{r}
\end{equation}
where $\delta=\sqrt{1-2\prob_0}=\bigo (\sqrt{n/2^n})$ and $\ket{r}$ is a residual normalized vector orthogonal to $\ket{\eigpert_0}$ and $\ket{-\eigpert_0}$. Now
\begin{equation}\label{Equation::Rotation}
\begin{array}{rcl}
\left(\upert\right)^\time\ket{\inita} &=&
\sqrt{\prob_0} (e^{i \eigpert_0 t} \ket{\eigpert_0}+e^{-i \eigpert_0 t} \ket{-\eigpert_0})+\delta \ket{r'}
 \\
&=& 2 \prob_0 \cos \eigpert_0 t \ket{\inita}-2\sqrt{\prob_0 \prob_1} (\sin \eigpert_0 t +\real e^{i \eigpert_0 t} \Delta) \ket{\initb}\\
&+& \sqrt{1-\prob_0-\prob_1}(e^{i \eigpert_0 t} \ket{r_0}+e^{-i \eigpert_0 t} \ket{r_0^*})+\delta \ket{r}\\
&=& \cos{\eigpert_0\time}\ket{\inita}  -\sin{\eigpert_0\time}\ket{\initb} + \bigo (\frac{n^{3/4}}{\sqrt{2^n}})\ket{\tilde{r}}
\end{array}
\end{equation}
where $\ket{\tilde{r}}$ is some residual normalized vector 
(not necessarily orthogonal to $\ket{\inita}$ and $\ket{\initb}$)
.

Starting with $\ket{\inita}$ and applying $\upert$ for  $t_f=\frac{\pi}{2\abs{\eigpert_0}}$ steps, we approximately rotate from $\ket{\inita}$ to $\ket{\initb}$. From $\ket{\initb}$ we can obtain $\ket{\vec{x}_{target}}=\ket{\vec{0}}$ with high probability $p$, since from \refeqn{InitbCollapsed} and with $1+1/2n \leq c \leq 1+2/\numbit$ for large $n$
\begin{equation}
\begin{array}{rcl}
\prob &=& \frac{1}{c}\sum_\dhc{\abs{\bkm{\dhc,0}{\initb}}^2}
= \frac{1}{c}\abs{\bkm{R,0}{\initb}}^2 \\
&\geq& \frac{1/2}{1+2/\numbit} = \frac{1}{2} - \bigo\left(1/\numbit\right)
\end{array}
\end{equation}
Finally, to obtain $t_f$ in terms of $\numbit$, we make use of the bounds on $\eigpert_0$ provided by \refthm{EigImpGap}: 
\begin{eqnarray}
t_f &=&\frac{\pi c}{2} \sqrt{2^{n-1}}(1 \pm \bigo(\frac{n^{3/2}}{\sqrt{2^n}})) \nonumber\\
&=& \frac{\pi}{2} \sqrt{2^{n-1}}(1 + \bigo (\frac{1}{n}))
\end{eqnarray}
If we set the number of time steps to be $t_f=\frac{\pi}{2}\sqrt{2^{n-1}}$ (or the closest integer) then 
\begin{equation}
-\sin \eigpert_0 t_f = \sin \frac{\pi}{2}(1-\bigo(\frac{1}{n})) = 1-\bigo(\frac{1}{n^2}).
\end{equation}
So the probability to measure $\ket{\vec{x}_{target}}$ after $t_f=\frac{\pi}{2}\sqrt{2^{n-1}}$ steps is still $p_{success}=1/2-O(1/n)$. 
Hence, by repeating the algorithm a constant number of times, the
probability of error can be made arbitrarily small. Note  the periodic nature of the evolution under $\upert$ (\refeqn{Rotation}); this means that if we measure at $t > t_f$ the probability of success will decrease and later increase again.  

In summary we arrived at the final result that the marked state is identified after $O(\sqrt{N})$ calls to the oracle.

\section{Connection to Grover's Algorithm} \label{Section::Grover}
The operation of the random walk search algorithm is similar in many
ways to the operation of Grover's search algorithm.  Both algorithms
begin in the equal superposition state over all bit strings.  Both
algorithms make use of the Grover diffusion operator, $\grover$,
(sometimes known as the Grover iterate).  Both algorithms can be
viewed as a rotation in a two-dimensional subspace.  Both algorithms
use an oracle which marks the target state with a phase of $-1$.  Both
algorithms have a running time of $\bigo\left(\sqrt{\numdim}\right)$. In both algorithms we have to measure at a specific time to obtain maximum probability of success. 
However, there are several important differences between the two
search algorithms.  In this section, we call attention to the
ways in which the random walk search algorithm is distinct from
Grover's algorithm and consider how these differences affect
performance and implementation.

It is well-known that Grover's algorithm can be mapped exactly onto a
rotation in the two-dimensional subspace spanned by the
equal-superposition state $\ket{\inita}$ and the marked state
$\ket{0}$ \cite{Grover:96}.  Each iteration in Grover's algorithm
corresponds to a rotation in this subspace.  In this paper, we have
shown that the random walk search algorithm can also be viewed as a
rotation in a two-dimensional subspace.  However, there are two
important distinctions.  First, the random walk search algorithm can
only be \emph{approximately} mapped onto a two-dimensional subspace.
Unlike Grover's algorithm, this mapping is not exact.  Second, the
two-dimensional subspace in which the random walk search algorithm is
approximately contained is spanned by $\ket{\inita}$ and
$\ket{\initb}$, not by $\ket{\inita}$ and $\ket{0}$.  Hence, the final
state of the algorithm is not exactly the pure marked state,
$\ket{0}$, as it is in Grover's algorithm.  It is a linear combination
of states which is composed primarily of the marked state, but also
possesses small contributions from its nearest neighbors,
second-nearest neighbors, etc.  Thus, the random walk search algorithm
contains traces of the underlying topology of the hypercube on which
it is based.

Another difference between the two algorithms is their use of the
Grover diffusion operator, $G$.  In Grover's algorithm, this operator
is applied to the entire $2^\numbit$-dimensional search space
(corresponding to the node space in the random walk search algorithm).
On the other hand, Grover's diffusion operator, $G$, in the random
walk algorithm is used as the quantum coin, and acts only on the
$\numbit$-dimensional coin space.  This fact may be of practical use
for certain physical implementations since many physical
implementations of quantum computers contain multiple types of qubits,
which have different natural gate sets.  We could exploit this variety
using the random walk search algorithm by choosing the coin space to
be represented by qubits on which it is convenient to implement the
Grover diffusion operator.

Another similarity between the two algorithms is the implementation of
the oracle.  In Grover's algorithm, the oracle marks the target state
with a phase of $-1$.  To arrive at this random walk search algorithm,
we chose the marking coin $\coin_1$ to be the $-\id$ coin.  This
choice was actually motivated because it yielded a result that was
amenable to analysis, and while the emergence of Grover's algorithm
appears natural in hindsight, it was not obvious at the outset.
However, more generally, it is not clear whether this choice of marked
coin is either optimal or unique.  In fact, numerical simulations have
shown us that many different types of marking coins will yield search
algorithms.  Unfortunately, analytic treatment of the quantum random
walk for more complicated coins has proven substantially more
difficult than the instance analyzed here for $\coin_1 = -\id$.  It is
an open question what (constant factor) gains might be made by using
different marking coins to implement the search.

\section{Conclusions} \label{Section::Conclusions}
In this paper, we have shown that the random walk search algorithm can
search a list of $2^{\numbit}$ items in time proportional to
$\sqrt{2^{\numbit}}$.  The lower bound on a quantum search of an
$\numdim$-item list is known to be
$\Omega\left(\sqrt{\numdim}\right)$\cite{Bennett:97}.  Thus, up to a
constant factor, the random walk search algorithm is optimal.
However, although after repetition of the algorithm a constant number
of times the result is arbitrarily close to the result of Grover's
search, the random walk search algorithm is not identically equivalent
to Grover's algorithm.  In particular, the final solution obtained by
the random walk search still retains some of the underlying character
of the hypercube on which it was based, with a small admixture of
states other than the solution at the marked node.

The random walk search analyzed here was based on a discrete walk on
the hypercube.  In general a similar methodology can be applied to any
regular graph, {\it e.g.}, a two-dimensional hexagonal lattice with
periodic boundary conditions, a three-dimensional rectangular lattice
with periodic boundary conditions, etc.  We have numerical evidence
indicating that this methodology will yield quantum search algorithms
when applied to other regular $\numbit$-dimensional lattices.  Future
studies will investigate the extent of optimality of such search
algorithms.

The intriguing possibility of finding novel algorithms based on the
random walk also remains an open question.  The results described here
indicate that the random walk search algorithm provides a suggestive
framework for new algorithms.  Though the optimality of Grover's
algorithm precludes the construction of an improved oracle-based
search algorithm based on a quantum walk, nevertheless, many other
oracle problems still exist for which a quantum walk may be
advantageous.  For instance, the lower bound on quantum search holds
only for oracles which provide ``yes/no''
information~\cite{Bennett:97}.  Our choice of marking coin here has a
clear relation to an identifiable component of Grover's algorithm.  In
general, the marking coin can be an arbitrary $\numbit\by\numbit$
unitary matrix.  The marking coin provides a intuitive means by which
to introduce a large amount of information to an oracle problem.
Thus, it is possible that unique coins with interesting properties may
give rise to an entirely new algorithm.  Overall we conclude that the
quantum random walk provides a means for insight into existing quantum
algorithms and offers a potentially vast source for development of new
algorithms.

\pagebreak[4]
\begin{figure}
\includegraphics{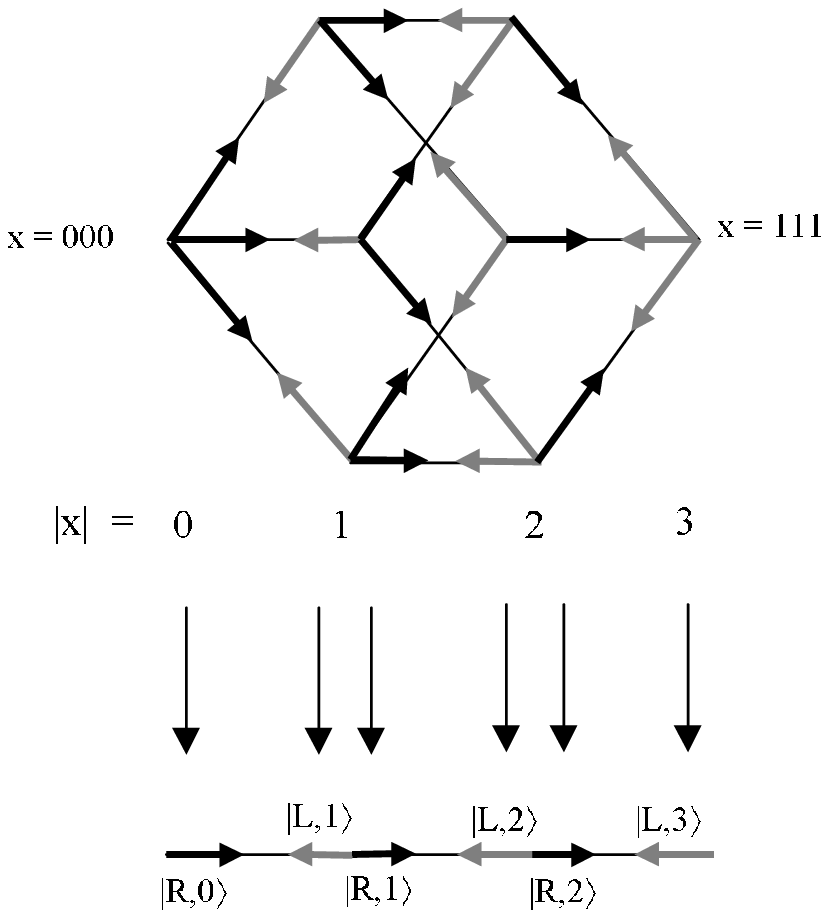}
\caption{Collapsing a random walk on the hypercube to a
random
walk on
the line.  States on the hypercube are mapped to state on the line
based on their
Hamming weight and the direction in which they
point (see text).}
\label{Figure::CubeToLine}
\end{figure}

\pagebreak[4]

\begin{figure}
\includegraphics{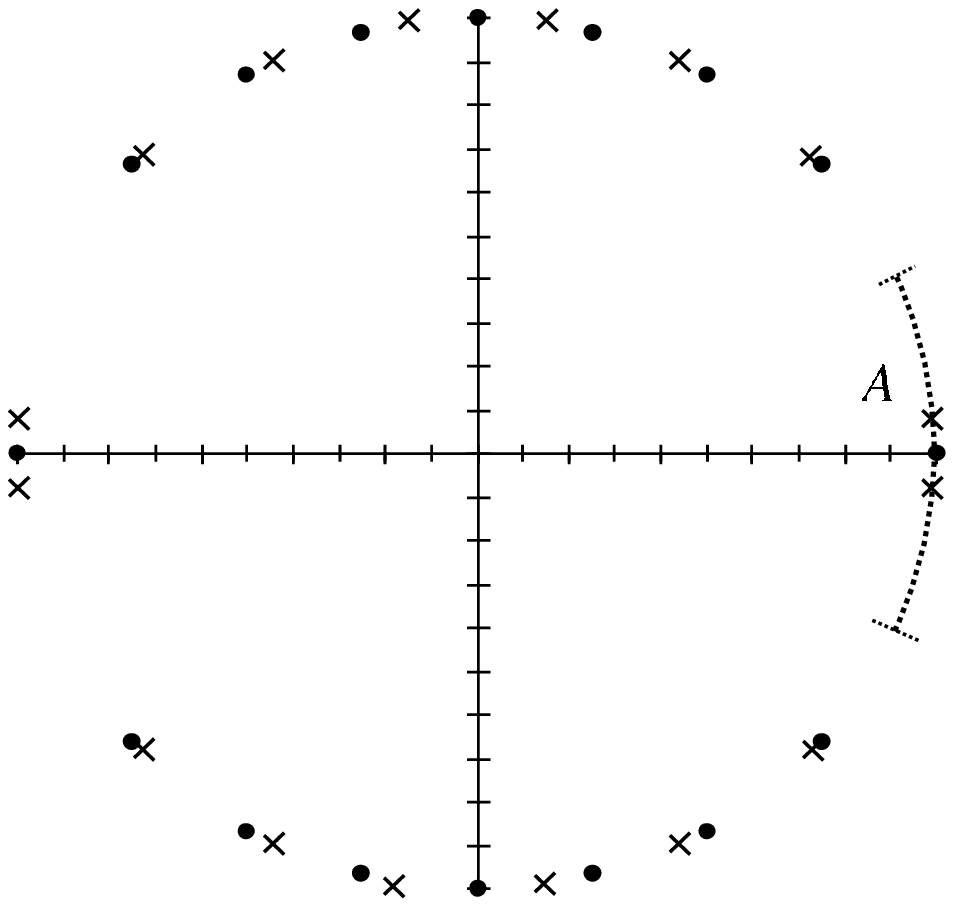}
\caption{The results of numerical spectral analysis of $\uunpert$
and
$\upert$ for $\numbit=8$.  Circles indicate eigenvalues of
$\uunpert$.
Crosses indicate eigenvalues of
$\upert$.}\label{Figure::NumericalSpectra}
\end{figure}

\pagebreak[4]
\begin{figure}
\includegraphics{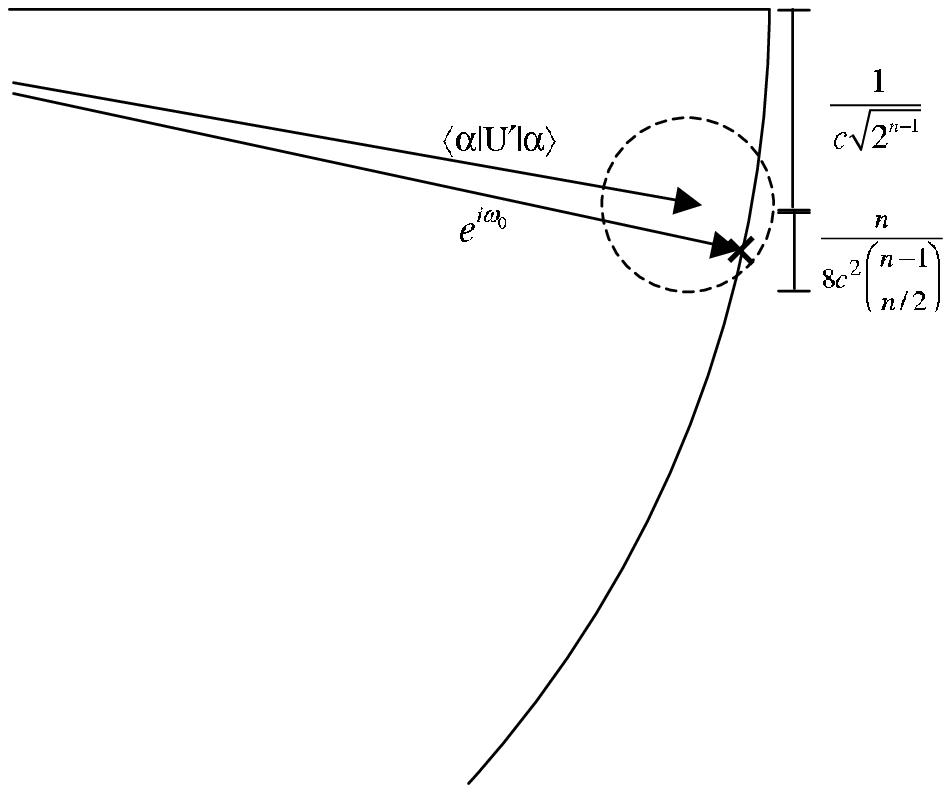}
\caption{Geometric representation of \refthm{EigImpGap}, which
proves
that the eigenvalue, $\e^{i\eigpert_0}$, must be located on a disc
of
radius $\frac{\numbit}{8\norminitb^2\nchoosek{\numbit-
1}{\numbit/2}}$
centered at $\bra{\eigtrial}\upert\ket{\eigtrial}$.  The position
of the
eigenvalue is denoted by a cross.}
\label{Figure::EigImpGap} \end{figure}

\begin{acknowledgments}
NS thanks the University of California, Berkeley, for a 
Berkeley Fellowship. This effort is sponsored by the Defense Advanced
Research Projects Agency (DARPA) and the Air Force Laboratory, Air
Force Material Command, USAF, under agreement number F30602-01-2-
0524. We also thank NSF ITR/SY award 0121555.

\end{acknowledgments}


\begin{thebibliography}{14}
\expandafter\ifx\csname natexlab\endcsname\relax\def\natexlab#1{#1}\fi
\expandafter\ifx\csname bibnamefont\endcsname\relax
  \def\bibnamefont#1{#1}\fi
\expandafter\ifx\csname bibfnamefont\endcsname\relax
  \def\bibfnamefont#1{#1}\fi
\expandafter\ifx\csname citenamefont\endcsname\relax
  \def\citenamefont#1{#1}\fi
\expandafter\ifx\csname url\endcsname\relax
  \def\url#1{\texttt{#1}}\fi
\expandafter\ifx\csname urlprefix\endcsname\relax\def\urlprefix{URL }\fi
\providecommand{\bibinfo}[2]{#2}
\providecommand{\eprint}[2][]{\url{#2}}

\bibitem[{\citenamefont{Aharonov et~al.}(1993)\citenamefont{Aharonov,
  Davidovich, and Zagury}}]{Aharonov:93a}
\bibinfo{author}{\bibfnamefont{Y.}~\bibnamefont{Aharonov}},
  \bibinfo{author}{\bibfnamefont{L.}~\bibnamefont{Davidovich}},
  \bibnamefont{and} \bibinfo{author}{\bibfnamefont{N.}~\bibnamefont{Zagury}},
  \bibinfo{journal}{Phys. Rev. A} \textbf{\bibinfo{volume}{48}},
  \bibinfo{pages}{1687} (\bibinfo{year}{1993}).

\bibitem[{\citenamefont{Aharonov et~al.}(2001)\citenamefont{Aharonov, Ambainis,
  Kempe, and Vazirani}}]{Aharonov:01}
\bibinfo{author}{\bibfnamefont{D.}~\bibnamefont{Aharonov}},
  \bibinfo{author}{\bibfnamefont{A.}~\bibnamefont{Ambainis}},
  \bibinfo{author}{\bibfnamefont{J.}~\bibnamefont{Kempe}}, \bibnamefont{and}
  \bibinfo{author}{\bibfnamefont{U.}~\bibnamefont{Vazirani}}, in
  \emph{\bibinfo{booktitle}{Proceedings of ACM Symposium on Theory of Computing
  (STOC)}} (\bibinfo{publisher}{ACM}, \bibinfo{address}{New York, NY},
  \bibinfo{year}{2001}), pp. \bibinfo{pages}{50--59},
  \bibinfo{note}{\uppercase{L}ANL preprint quant-ph/0012090}.

\bibitem[{\citenamefont{Ambainis et~al.}(2001)\citenamefont{Ambainis, Back,
  Nayak, Vishwanath, and Watrous}}]{Ambainis:01}
\bibinfo{author}{\bibfnamefont{A.}~\bibnamefont{Ambainis}},
  \bibinfo{author}{\bibfnamefont{E.}~\bibnamefont{Back}},
  \bibinfo{author}{\bibfnamefont{A.}~\bibnamefont{Nayak}},
  \bibinfo{author}{\bibfnamefont{A.}~\bibnamefont{Vishwanath}},
  \bibnamefont{and} \bibinfo{author}{\bibfnamefont{J.}~\bibnamefont{Watrous}},
  in \emph{\bibinfo{booktitle}{Proc. 33rd STOC}} (\bibinfo{publisher}{ACM},
  \bibinfo{address}{New York, NY}, \bibinfo{year}{2001}), pp.
  \bibinfo{pages}{60--69}.

\bibitem[{\citenamefont{Farhi and Gutmann}(1998)}]{Farhi:97}
\bibinfo{author}{\bibfnamefont{E.}~\bibnamefont{Farhi}} \bibnamefont{and}
  \bibinfo{author}{\bibfnamefont{S.}~\bibnamefont{Gutmann}},
  \bibinfo{journal}{Phys. Rev. A} \textbf{\bibinfo{volume}{58}},
  \bibinfo{pages}{915} (\bibinfo{year}{1998}).

\bibitem[{\citenamefont{Childs et~al.}(2001)\citenamefont{Childs, Farhi, and
  Gutmann}}]{Childs:01}
\bibinfo{author}{\bibfnamefont{A.}~\bibnamefont{Childs}},
  \bibinfo{author}{\bibfnamefont{E.}~\bibnamefont{Farhi}}, \bibnamefont{and}
  \bibinfo{author}{\bibfnamefont{S.}~\bibnamefont{Gutmann}},
  \bibinfo{journal}{quant-ph/0103020}  (\bibinfo{year}{2001}).

\bibitem[{\citenamefont{Childs et~al.}(2002)\citenamefont{Childs, Cleve,
  Deotto, Farhi, Gutmann, and Spielman}}]{Childs:02}
\bibinfo{author}{\bibfnamefont{A.}~\bibnamefont{Childs}},
  \bibinfo{author}{\bibfnamefont{R.}~\bibnamefont{Cleve}},
  \bibinfo{author}{\bibfnamefont{E.}~\bibnamefont{Deotto}},
  \bibinfo{author}{\bibfnamefont{E.}~\bibnamefont{Farhi}},
  \bibinfo{author}{\bibfnamefont{S.}~\bibnamefont{Gutmann}}, \bibnamefont{and}
  \bibinfo{author}{\bibfnamefont{D.}~\bibnamefont{Spielman}},
  \bibinfo{journal}{quant-ph/0209131}  (\bibinfo{year}{2002}).

\bibitem[{\citenamefont{Grover}(1996)}]{Grover:96}
\bibinfo{author}{\bibfnamefont{L.}~\bibnamefont{Grover}}, in
  \emph{\bibinfo{booktitle}{Proc. $28^{th}$ Annual ACM Symposium on the Theory
  of Computation}} (\bibinfo{publisher}{ACM Press}, \bibinfo{address}{New York,
  NY}, \bibinfo{year}{1996}), pp. \bibinfo{pages}{212--219}.

\bibitem[{\citenamefont{Grover}(1997)}]{Grover:97}
\bibinfo{author}{\bibfnamefont{L.}~\bibnamefont{Grover}},
  \bibinfo{journal}{Phys. Rev. Lett.} \textbf{\bibinfo{volume}{79}},
  \bibinfo{pages}{325} (\bibinfo{year}{1997}).

\bibitem[{\citenamefont{Bennett et~al.}(1997)\citenamefont{Bennett, Bernstein,
  Brassard, and Vazirani}}]{Bennett:97}
\bibinfo{author}{\bibfnamefont{C.}~\bibnamefont{Bennett}},
  \bibinfo{author}{\bibfnamefont{E.}~\bibnamefont{Bernstein}},
  \bibinfo{author}{\bibfnamefont{G.}~\bibnamefont{Brassard}}, \bibnamefont{and}
  \bibinfo{author}{\bibfnamefont{U.}~\bibnamefont{Vazirani}},
  \bibinfo{journal}{SIAM J. Comput.} \textbf{\bibinfo{volume}{26}},
  \bibinfo{pages}{1510} (\bibinfo{year}{1997}).

\bibitem[{\citenamefont{Nielsen and Chuang}(2000)}]{Nielsen}
\bibinfo{author}{\bibfnamefont{M.}~\bibnamefont{Nielsen}} \bibnamefont{and}
  \bibinfo{author}{\bibfnamefont{I.}~\bibnamefont{Chuang}},
  \emph{\bibinfo{title}{Quantum Computation and Quantum Information}}
  (\bibinfo{publisher}{Cambridge University Press},
  \bibinfo{address}{Cambridge}, \bibinfo{year}{2000}).

\bibitem[{\citenamefont{Moore and Russell}(2002)}]{Moore:01}
\bibinfo{author}{\bibfnamefont{C.}~\bibnamefont{Moore}} \bibnamefont{and}
  \bibinfo{author}{\bibfnamefont{A.}~\bibnamefont{Russell}},
  \bibinfo{journal}{Proc. RANDOM, to appear}  (\bibinfo{year}{2002}),
  \bibinfo{note}{lanl-ar{X}ive quant-ph/0104137}.

\bibitem[{\citenamefont{Kempe}(2002)}]{Kempe:02}
\bibinfo{author}{\bibfnamefont{J.}~\bibnamefont{Kempe}},
  \bibinfo{journal}{quant-ph/0205083}  (\bibinfo{year}{2002}).

\bibitem[{\citenamefont{Yamasaki et~al.}(2002)\citenamefont{Yamasaki,
  Kobayashi, and Imai}}]{Yamasaki:02}
\bibinfo{author}{\bibfnamefont{T.}~\bibnamefont{Yamasaki}},
  \bibinfo{author}{\bibfnamefont{H.}~\bibnamefont{Kobayashi}},
  \bibnamefont{and} \bibinfo{author}{\bibfnamefont{H.}~\bibnamefont{Imai}},
  \bibinfo{journal}{quant-ph/0205045}  (\bibinfo{year}{2002}).

\bibitem[{\citenamefont{Watrous}(2001)}]{Watrous:01a}
\bibinfo{author}{\bibfnamefont{J.}~\bibnamefont{Watrous}},
  \bibinfo{journal}{Journal of Computer and System Sciences}
  \textbf{\bibinfo{volume}{62}}, \bibinfo{pages}{376} (\bibinfo{year}{2001}).

\end{thebibliography}
\bibliographystyle{apsrev}

\end{document}